\documentclass[twocolumn,english]{revtex4}
\usepackage[T1]{fontenc}
\usepackage[utf8]{luainputenc}
\usepackage{amsmath}
\usepackage{amssymb}
\usepackage{graphicx}
\usepackage{esint}

\makeatletter
\@ifundefined{textcolor}{}
{%
 \definecolor{BLACK}{gray}{0}
 \definecolor{WHITE}{gray}{1}
 \definecolor{RED}{rgb}{1,0,0}
 \definecolor{GREEN}{rgb}{0,1,0}
 \definecolor{BLUE}{rgb}{0,0,1}
 \definecolor{CYAN}{cmyk}{1,0,0,0}
 \definecolor{MAGENTA}{cmyk}{0,1,0,0}
 \definecolor{YELLOW}{cmyk}{0,0,1,0}
}

\usepackage{babel}

\usepackage{babel}

\usepackage{babel}

\usepackage{babel}

\usepackage{babel}

\usepackage{babel}

\usepackage{babel}

\usepackage{babel}

\usepackage{babel}

\usepackage{babel}

\usepackage{babel}

\makeatother

\usepackage{babel}
\begin{document}

\title{Failure of the local GGE for integrable models with bound states}

\author{Garry Goldstein and Natan Andrei}

\address{Department of Physics, Rutgers University}

\address{Piscataway, New Jersey 08854}
\begin{abstract}
In this work we study the applicability of the local GGE to integrable
one dimensional systems with bound states. We find that the GGE, when
defined using only local conserved quantities, fails to describe the
long time dynamics for most initial states including eigenstates.
We present our calculations studying the attractive Lieb-Liniger gas
and the XXZ magnet, though similar results may be obtained for other
models. 
\end{abstract}
\maketitle

\section{\label{sec:Introduction}Introduction}

Recent years have witnessed spectacular advances in the theory of
unitary nonequilibrium dynamics, particularly in systems of optically
trapped atomic gases. Key to this advance is the extremely weak coupling
to the environment which allows for essentially Hamiltonian dynamics.
These experimental advances have spurred many theoretical questions:
does a steady state emerge, how do local observables equilibrate,
is there any principle which allows us to relate the steady state
to the initial conditions?

One of the most surprising recent experimental \cite{key-5}, and
theoretical \cite{key-10} results is that there is a relation between
the initial state and the long time steady state for integrable models.
It was shown that after a quench integrable models retain memory of
their initial state and do not appear to relax to thermodynamic equilibrium.
This was ascribed to the fact that integrable models possess an infinite
family of local conserved charges in involution, $\left\{ I_{i}\right\} $,
which include the Hamiltonian $H$, typically identified with $I_{2}$:
\begin{equation}
\left[H,I_{i}\right]=\left[I_{i},I_{i'}\right]=0,\, H=I_{2}\label{eq:Conserved_Charges}
\end{equation}

These conserved quantities in turn imply that there is a complete
system of eigenstates for the Hamiltonian which may be parametrized
by sets of rapidities $\left\{ k\right\} $ and which simultaneously
diagonalize all charges. To understand the equilibration of this system
it was recently proposed that it is insufficient to consider only
thermal ensembles but it is also necessary to include these nontrivial
conserved quantities. It was proposed \cite{key-34} that the system
relaxes to a state given by the generalized Gibbs ensemble GGE with
its density matrix being given by 
\begin{equation}
\rho_{GGE}=\frac{1}{Z}\exp\left(-\sum\alpha_{i}I_{i}\right)\label{eq:GGE_density_matrix}
\end{equation}
where the $I_{i}$ are the local conserved quantities; the $\alpha_{i}$
are the generalized inverse temperatures and $Z$ is a normalization
constant insuring $Tr\left[\rho_{GGE}\right]=1$. The $\alpha_{i}$
are chosen in such a way as to insure that the conserved quantities
$I_{i}$ remain constant, namely, $\langle I_{i}\rangle_{final}\equiv Tr\left\{ \rho_{GGE}I_{i}\right\} =\left\langle I_{i}\left(t=0\right)\right\rangle \equiv\langle I_{i}\rangle_{initial}=I_{i}^{0}$.
Moreover it was proposed that expectation values of local operators
and of correlation functions of an integrable model may be computed
at long times by taking their expectation value with respect to the
GGE density matrix, e.g. $\left\langle \Theta\left(t\rightarrow\infty\right)\right\rangle =Tr\left[\rho_{GGE}\Theta\right]$.
Recent numeric and theoretical works have, however, put this assumption
into question \cite{key-32}.

Here we would like to show that the GGE hypothesis, based on local
conserved quantities, fails in general for the class of integrable
models possessing bound states, or string eigenstates. Bound states
in integrable models are described, in the thermodynamic limit, by
rapidities forming $n$-strings, \cite{key-31}: $k_{\alpha}^{j}=k_{\alpha}+i\mu(n-2j),\; j=0,1,2,\dots n$,
with $n$ an arbitrary integer and $\mu$ a coupling constant in the
Hamiltonian. We will show that for such models the GGE hypothesis
fails to reproduce the long time dynamics for most states and in particular
for eigenstates of the Hamiltonian. We will focus in detail on the
attractive Lieb-Liniger model, and repeat our arguments more briefly
for the XXZ model. Our results are also applicable to other models
with bound states.

The Lieb Liniger hamiltonian is given by: 
\begin{equation}
H_{LL}=\intop_{-\infty}^{\infty}dx\left\{ \partial_{x}b^{\dagger}\left(x\right)\partial_{x}b\left(x\right)+c\left(b^{\dagger}\left(x\right)b\left(x\right)\right)^{2}\right\} ,\label{eq:lieb-lin-hamiltonian}
\end{equation}
Here $b^{\dagger}\left(x\right)$ is the bosonic creation operator
at the point $x$ and $c$ is the coupling constant. The eigenstates
of the Hamiltonian are parametrized by rapidities $\left|\left\{ k\right\} \right\rangle $.
In the basis of the Bethe rapidities, $\left|\left\{ k\right\} \right\rangle $,
the local conserved quantities $I_{i}$ are diagonalized and take
the form, $I_{i}\left|\left\{ k\right\} \right\rangle =\sum k^{i}\left|\left\{ k\right\} \right\rangle $.
It was pointed out recently \cite{key-3} that when not acting on
eigenstates (or on a finite linear combination of them) the charges
may generate divergences in the form of powers and derivatives of
Dirac-deltas. Great care must then be taken to define their action.
Here we assume an appropriate renormalization scheme has been implemented
\cite{key-4}.

We consider the attractive Hamiltonian with the coupling constant
taken to be negative $c<0$. In this case bound states are formed
and the rapidities fall into $n$-string configurations $k_{j}=k_{0}+\frac{ic}{2}\left(n-2j\right)$,
with strings of arbitrary length length $n=1,2,3\cdots$ The contribution
an $n$-string centered at $k_{0}$ to the conserved charge $I_{i}$
is given by: 
\begin{equation}
\begin{array}{l}
\varepsilon_{i}^{n}\left(k_{0}\right)\equiv\sum_{j=0}^{n}\left(k_{0}+\frac{ic}{2}\left(n-2j\right)\right)^{i}=\\
=\sum_{l=0}^{i}k_{0}^{i-l}\left(\begin{array}{c}
i\\
l
\end{array}\right)\left(\frac{ic}{2}\right)^{i}\sum_{j=1}^{n}\left(n-2j\right)^{i}
\end{array}\label{eq:Conserved_Quantity}
\end{equation}
We will show here explicitly for the Lieb-Liniger and the XXZ models
(the proof can be extended to other models with bound states such
as the Hubbard model or the Anderson model) that when the system is
quenched from a non-equilibrium initial state $\left|\Phi_{0}\right\rangle $
and allowed to evolve for a long time, the GGE hypothesis fails and
for most initial states $\left|\Phi_{0}\right\rangle $, including
eigenstates, does not provide a correct description of the equilibrated
system.

The rest of the paper is organized as follows. In Section \ref{sec:Outline_Proof}
we present the outline of the proof of this result; in subsection
\ref{sec:Properties_State} we present some properties of the conserved
quantities of the states used in Section \ref{sec:Outline_Proof};
in subsection \ref{sec:Properties_GGE} we present some results concerning
the GGE used in Section \ref{sec:Outline_Proof}, in Section \ref{sec:XXZ-Model}
we briefly discuss the XXZ model and in Section \ref{sec:Conclusions}
we conclude.

\section{\label{sec:Outline_Proof}Outline of proof}

For the attractive Lieb-Liniger gas we may introduce densities of
string excitations; for a given eigenstate $\left|\left\{ k\right\} \right\rangle $
we denote by $\rho_{p}^{n}\left(k\right)$ the Bethe density of $n$-strings,
so that $L\rho_{p}^{n}\left(k\right)dk$ is the number of $n$-strings
in the interval $\left[k,k+dk\right]$. Similarly $\rho_{h}^{n}\left(k\right)$
denotes the $n$-strings hole density and $\rho_{t}^{n}\left(k\right)=\rho_{p}^{n}\left(k\right)+\rho_{h}^{n}\left(k\right)$
the total $n$-string density. The Yang-Yang entropy associated with
the densities, $\{\rho_{p}^{n}\left(k\right),\rho_{h}^{n}\left(k\right)\}$,
measures the number of states $\left|\left\{ k\right\} \right\rangle $
consistent with the densities. It is given by: 
\begin{equation}
\begin{array}{l}
S\left(\left\{ \rho^{n}\right\} \right)=\\
=\sum_{n=0}^{\infty}\int_{-\infty}^{\infty}dk\left(\rho_{h}^{n}\left(k\right)\ln\left(\frac{\rho_{t}^{n}\left(k\right)}{\rho_{h}^{n}\left(k\right)}\right)+\rho_{p}^{n}\left(k\right)\ln\left(\frac{\rho_{t}^{n}\left(k\right)}{\rho_{p}^{n}\left(k\right)}\right)\right).
\end{array}\label{eq:Entorpy}
\end{equation}
With this notation our proof of the failure of the GGE, based on local
conserved quantities, will be based on two main results which we prove
below: (1) For the attractive Lieb Liniger model for a given set of
conserved quantities $I_{i}^{0}$ there is an infinite number of densities
$\left\{ \rho_{p}^{n}\right\} $ satisfying $I_{i}\left\{ \rho_{p}^{n}\right\} =I_{i}^{0}$,
(2) The GGE corresponds to a pure state whose Bethe densities $\left\{ \rho^{n}\right\} $
maximizes the entropy $S\left(\left\{ \rho^{n}\right\} \right)$ subject
to the thermodynamic Bethe Ansatz equation and the constraints $I_{i}\left\{ \rho_{p}^{n}\right\} =I_{i}^{0}$.

\begin{figure}
\begin{centering}
\includegraphics[width=1\columnwidth]{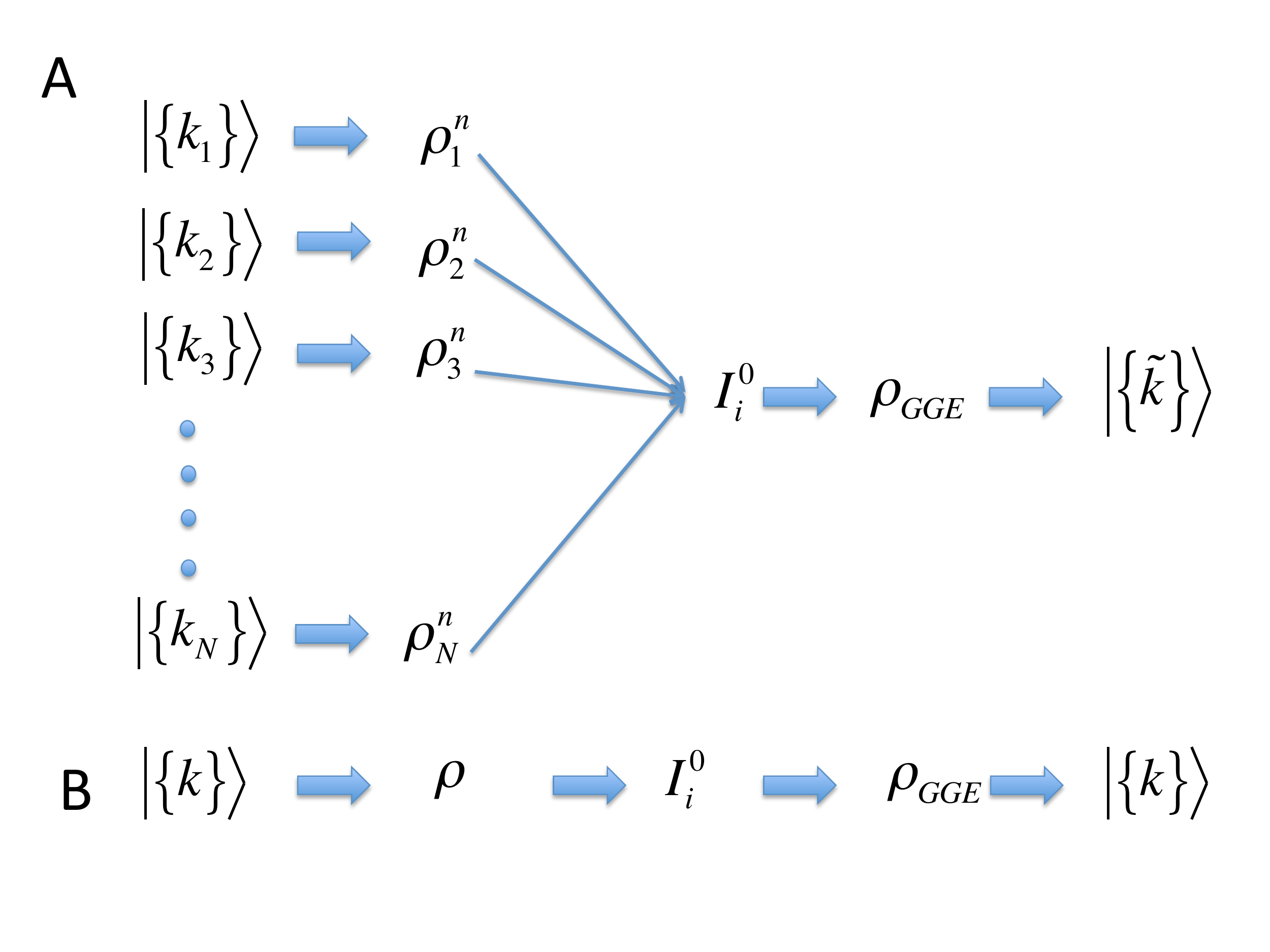} 
\par\end{centering}

\protect\protect\caption{\label{fig:GGE_Logic}The logic of the GGE argument. A) Attractive
Lieb-Liniger model. There are many exact eigenstates whose particle
densities are different but which correspond to the same conserved
quantities $I_{i}^{0}$. The $I_{i}^{0}$ however determine a unique
GGE density matrix which corresponds to a specific pure state. B)
Repulsive Lieb-Liniger model. For each set of conserved quantities
there is only one quasiparticle density which corresponds to one GGE
density matrix which is then equivalent to the original quasiparticle
density.}
\end{figure}

It follows then that while in the repulsive Lieb-Liniger model the
quantities $I_{i}^{0}$, fixed by the initial state, uniquely determine
the Bethe density $\rho_{p}$ such that $I_{i}\{\rho_{p}\}=I_{i}^{0}$,
no such shortcut is available in the case of the attractive model.
An infinite number of densities are required its description and the
conditions $I_{i}\left\{ \rho_{p}^{n}\right\} =I_{i}^{0}$ are insufficient
to determine the densities and hence the GGE. The full time evolution
is required. In more detail, point (1) indicates that there are many
states with different particle densities $\left\{ \rho_{p}^{n}\right\} $
that give the same GGE. However states with different densities have
different local correlations \cite{key-2} so cannot correspond to
the same density matrix. This is sufficient to show the failure of
the GGE for most eigenstates, see Fig. (\ref{fig:GGE_Logic}). The
point (2) identifies which eigenstates correspond to the GGE. For
these eigenstates and these eigenstates only the GGE is a good description
of the state.

If furthermore we assume the o-TBA hypothesis \cite{key-32,key-33}
then most states correspond to eigenstates and we get that the GGE
fails for most states. We note that we do not need an o-TBA assumption
to show that GGE based on local conserved quantities fails.

In Section \ref{sec:Properties_State} we prove property (1) and in
Section \ref{sec:Properties_GGE} we prove property (2).

\subsection{\label{sec:Properties_State}Properties of the states}

We would like to show that for a given set of conserved quantities
$\left\{ I_{i}^{0}\right\} $ there is an infinite set of different
densities $\left\{ \rho_{p}^{n}\right\} $ that satisfy $I_{i}\left\{ \rho_{p}^{n}\right\} =I_{i}^{0}$.
In order to accomplish this we introduce the notation $J_{i}^{n}=\int dk\rho_{p}^{n}\left(k\right)k^{i}$.
These are the moments of the distributions $\rho_{p}^{n}$. There
is a one to one correspondence between a distribution and its moments.
Then using Eq. (\ref{eq:Conserved_Quantity}) the equation $I_{i}\left\{ \rho_{p}^{n}\right\} =I_{i}^{0}$
may be written as: 
\begin{equation}
\sum_{n=0}^{\infty}\sum_{l=0}^{i}J_{l}^{n}\left(\frac{ic}{2}\right)^{i-l}\sum_{j=0}^{n}\left(n-2j\right)^{i-l}=I_{i}^{0}.\label{eq:Density_conserved_quantities}
\end{equation}
We note that it is a linear equation in the quantities $J_{l}^{n}$.
In particular it can be written as $\sum_{n=0}^{\infty}M_{l}^{i}J_{l}^{n}=I_{i}^{0}$,
with $M_{l}^{i}=\theta\left(i-l\right)\left(\frac{ic}{2}\right)^{i-l}\sum_{j=0}^{n}\left(n-2j\right)^{i-l}$.
It is easy to see that if there is at least one solution there are
infinitely many solutions. Indeed, this a a vastly underdetermined
set of linear equations. There are infinitely more variables then
constraints. Any element in the kernel of the transformation in Eq.
(\ref{eq:Density_conserved_quantities}), which is infinite dimensional,
may be added to any solution to obtain another solution. To be completely
explicit assume $\left\{ \rho_{p}^{n0}\right\} $ is a solution which
corresponds to a set of moments of the form $\left\{ J_{l}^{n0}\right\} $.
Then choose an $n$ such that $J_{l}^{n0}$ is not zero and is in
the interior of the set of allowed moments. Now choose an arbitrary
small deformation of the rest of the moments $J_{l}^{m0}+\delta J_{l}^{m}$
$m\neq n$ such that the moments $J_{l}^{m0}+\delta J_{l}^{m}$ corresponds
to a real densities $\rho_{p}^{n}$. There is an infinite number of
ways to do this. Then in order for Eq. (\ref{eq:Density_conserved_quantities})
to be satisfied all we need is to choose a change in the last moment
$\delta J_{l}^{n}=-\left(M^{-1}\right)_{l}^{i}\sum_{m\neq n}\sum_{l=0}^{i}\delta J_{l}^{n}\left(\frac{ic}{2}\right)^{i-l}\sum_{j=0}^{m}\left(m-2j\right)^{i-l}$.
We note that $M_{l}^{i}$ is lower triangular with all the diagonal
entries equal to $n+1$ hence invertible. Since $J_{l}^{m=0}$ is
not zero and is in the interior of the set of allowed moments and
$\delta J_{l}^{n}$ is small we have the moments $J_{l}^{m=0}+\delta J_{l}^{n}$
also correspond to a real density $\rho_{p}^{n}$. As such there is
an infinite number of densities corresponding to each set of conserved
moments.

\subsection{\label{sec:Properties_GGE}Properties of the GGE}

We now show that the GGE density matrix corresponding to the moments
$I_{i}^{0}$ corresponds to the pure state with the same moments that
maximizes the entropy given in Eq. (\ref{eq:Entorpy}). To do so we
use the result in \cite{key-1} that the GGE, based on local conserved
quantities, corresponds to a pure state that maximizes the functional
$\Xi\left(\left\{ \rho^{n}\right\} \right)\equiv-\sum\alpha_{i}\sum_{n=0}^{\infty}\int dk\rho_{p}^{n}\left(k\right)\varepsilon_{n}^{i}\left(k\right)+S\left(\rho^{n}\right)$,
subject to the constraint of the Thermodynamic Bethe Ansatz. The quantities
$\alpha_{i}$ are chosen such that this pure state satisfies $I_{i}\left(\left\{ \rho_{p}\left(k\right)\right\} \right)\equiv\sum_{n=0}^{\infty}\int dk\rho_{p}^{n}\left(k\right)\varepsilon_{n}^{i}\left(k\right)=I_{i}^{0}$
so the maximization happens when this is satisfied. Within this subspace
the functional $\Xi\left(\left\{ \rho^{n}\right\} \right)$ simplifies
to $\Xi\left(\left\{ \rho^{n}\right\} \right)=-\sum\alpha_{i}I_{i}^{0}+S\left(\left\{ \rho^{n}\right\} \right)$.
Therefore the GGE corresponds to the pure state that has the prescribed
conserved quantities and maximizes the quantity $S\left(\left\{ \rho^{n}\right\} \right)$
subject to the TBA and the constraint that $I_{i}\left(\left\{ \rho_{p}\left(k\right)\right\} \right)=I_{i}^{0}$.
Therefore for fixed conserved quantities the GGE corresponds to a
single pure state and only reproduces the long time dynamics of that
pure state.

\section{\label{sec:XXZ-Model}XXZ Model}

We will briefly discuss how to extend our results to the XXZ model.
The XXZ hamiltonian is given by: 
\begin{equation}
H_{XXZ}=-J\sum_{i=\infty}^{\infty}\left[\sigma_{i}^{x}\sigma_{i+1}^{x}+\sigma_{i}^{y}\sigma_{i+1}^{y}+\Delta\left(\sigma_{i}^{z}\sigma_{i+1}^{z}-1\right)\right]\label{eq:XXZ_Hamiltonian}
\end{equation}

Here $\sigma_{i}^{x,y,z}$ is the Pauli matrix at the site $i$, we
will focus on the case where $\Delta>1$. This Hamiltonian is integrable
and has an infinite number of conserved quantities $\tilde{I}_{i}$.
The eigenstates may be parametrized by rapidities $\left|\left\{ \lambda\right\} \right\rangle $.
The rapidities may be arranged into strings where each string is composed
of the rapidities given by $\lambda_{j}=\lambda+\frac{i\eta}{2}\left(n-2j\right)$
with $j=0,1,2,..n$ and $\lambda\in\left[-\frac{\pi}{2},\frac{\pi}{2}\right]$
and $\Delta=\cosh\eta$. We would like to repeat our proof of the
failure of the GGE for eigenstates for the case of the XXZ Hamiltonian.
To do so we need to prove two statements akin to what was done in
section (\ref{sec:Outline_Proof}): (1) That for a given set of conserved
quantities $\tilde{I}_{i}$ there is an infinite number of possible
quasiparticle densities $\left\{ \tilde{\rho}^{n}\right\} $, (2)
The GGE corresponds to a single choice of these densities $\tilde{\rho}^{n}$
which maximizes the entropy see Eq. (\ref{eq:Entorpy}). To prove
(2) we note that according to the result in \cite{key-1} that the
GGE corresponds to a pure state that maximizes the functional $\tilde{\Xi}\left(\left\{ \rho^{n}\right\} \right)\equiv-\sum\alpha_{i}\sum_{n=0}^{\infty}\int_{-\frac{\pi}{2}}^{\frac{\pi}{2}}dk\tilde{\rho}_{p}^{n}\left(\lambda\right)\tilde{\varepsilon}_{n}^{i}\left(\lambda\right)+S\left(\tilde{\rho}^{n}\right)$
subject to the constraint of the Thermodynamic Bethe Ansatz. Here
$\tilde{\varepsilon}_{n}^{i}$ is the value of the i'th conserved
quantities when acting on a string and $S\left(\tilde{\rho}^{n}\right)$
is completely analogous to Eq. (\ref{eq:Entorpy}). We know that the
quantities $\alpha_{i}$ are chosen such that this pure state satisfies
$\tilde{I}_{i}\left(\left\{ \tilde{\rho}_{p}\left(\lambda\right)\right\} \right)\equiv\sum_{n=0}^{\infty}\int d\lambda\rho_{p}^{n}\left(\lambda\right)\tilde{\varepsilon}_{n}^{i}\left(\lambda\right)=\tilde{I}_{i}^{0}$.
Within this subspace the functional $\tilde{\Xi}\left(\left\{ \rho^{n}\right\} \right)$
simplifies to $\tilde{\Xi}\left(\left\{ \tilde{\rho}^{n}\right\} \right)=-\sum\alpha_{i}I_{i}^{0}+S\left(\left\{ \tilde{\rho}^{n}\right\} \right)$.
Therefore the GGE corresponds to the pure state that has the right
conserved quantities and maximizes the quantity $S\left(\left\{ \tilde{\rho}^{n}\right\} \right)$
subject to the TBA. To prove the observation (1) we note that the
quantities $\tilde{\varepsilon}_{n}^{i}\left(\lambda\right)$ have
convergent power series in $\sin\left(2\lambda\right)$ and $\cos\left(2\lambda\right)$
\cite{key-32} and therefore may be expressed as a linear function
in the quantities $\left\{ \sin\left(2n\lambda\right)\right\} $ and
$\left\{ \cos\left(2n\lambda\right)\right\} $. This means that the
quantities $\tilde{I}_{i}\left(\tilde{\rho}^{n}\right)$ may be related
to the Fourier coefficients of $\tilde{\rho}^{n}$. To reproduce values
of the conserved quantities $I_{i}^{0}$ we obtain equations similar
to Eq. (\ref{eq:Density_conserved_quantities}) where the quantities
$J_{l}^{n}$ are replaced by the Fourier coefficients $F_{l}^{n}$
of the quantities $\tilde{\rho}^{n}$. These equations are once again
vastly underdetermined with infinitely more variables then constraints.
As such, in a completely analogous way as before we obtain that there
is infinite number of solutions to these equations and as a result
the GGE fails for pure states for the XXZ Hamiltonian.

\section{\label{sec:Conclusions}Conclusions}

We have shown that for integrable models with bound states the GGE
based on local conserved quantities does not represent the long time
dynamics of many states, in particular most eigenstates. This result
is based on two observations we have proven: (1) for any set of conserved
quantities $I_{i}^{0}$ there are many states $\left|\left\{ k\right\} \right\rangle $
such that $I_{i}\left(\left|\left\{ k\right\} \right\rangle \right)=I_{i}^{0}$,
(2) The GGE corresponds to a specific state with fixed $I_{i}^{0}$.
We have verified these statements explicitly for the case of the attractive
Lieb-Liniger gas and the XXZ magnet though similar verifications may
be done for other integrable models with bound states.

The question whether the local charges form a complete set is of great
interest. For the repulsive case they appear to be  complete, as the GGE with
local charges represents the long time limit of the model. However
the GGE with local charges fails to do so when the coupling constant
changes sign. Would new non-local charges be required for the attractive
Lieb-Liniger model? For the XXZ Heisenberg model such charges were
recently proposed \cite{key-9} to treat another aspect of completeness
in the context of the Mazur inequality. Whether they also repair the
GGE is an open question.

\textbf{Acknowledgments}: This research was supported by NSF grant
DMR 1006684 and Rutgers CMT fellowship. We would like to thank M.
Rigol and J.-S. Caux and Marton Kormos for useful discussions and
comments.

\section*{\label{part:Appendix:-Regularization-schemes}Appendix: Regularization
schemes}

We would like to describe how to regularize the GGE for the attractive
Lieb-Liniger model. Multiple regularizations are possible \cite{key-3}.
For example a regularization scheme for the GGE for the Lieb Liniger
gas may be the following. Consider the conserved quantities: 
\begin{equation}
\tilde{I}_{i}\left|\left\{ k\right\} \right\rangle =\sum k^{i}\exp\left(-\lambda k^{2}\right)\left|\left\{ k\right\} \right\rangle \label{eq:Conserved quantities}
\end{equation}

These are semi local in the sense that they have convergent powerseries
with the expansion coefficients being the the usual conserved quantities
$I_{i}\left|\left\{ k\right\} \right\rangle =\sum k^{i}\left|\left\{ k\right\} \right\rangle $.
Here $\lambda$ is a positive real number. We can define a GGE with
these conserved quantities, that is we write $\rho_{GGE}=\frac{1}{Z}\exp\left(-\sum\alpha_{i}I_{i}\right)$
such that $tr\left\{ \rho_{GGE}\tilde{I}_{i}\right\} =\tilde{I}_{i}\left(t=0\right)$.
The $\tilde{I}_{i}$ have finite expectation values for any $\lambda$.
Furthermore the matrix $\rho_{GGE}$ is independent of $\lambda$.
Indeed for any $\lambda$ the $\tilde{I}_{i}$ are linear combinations
of the $I_{i}$. Furthermore this linear transformation $\tilde{I}_{i}\rightarrow I_{i}$
is invertible (the matrix of the transformation is upper triangular
with all ones on the diagonal). By composing the transformation $\tilde{I}_{i}^{\lambda_{1}}\rightarrow\tilde{I}_{i}^{\lambda_{2}}$is
also invertible. Conservation of one set of charges for one $\lambda$
is completely equivalent to conservation of another set of charges
for a different $\lambda$. With this regularization it is in principle
possible to define the GGE for any state. Another regularization scheme
is to average the local density for the conserved charges over a small
interval, thereby obtaining a finite result, use these conserved quantities
to calculate the GGE and then take the limit where the averaging goes
to zero.


\begin{thebibliography}{10}
\bibitem[1]{key-5} T. Kinoshita, T. Wenger, and D. S. Weiss, Nature
(London) \textbf{440}, 900 (2006)

\bibitem[2]{key-10} Kai He and M. Rigol, Phys. Rev. A 87, 043615
(2013); G. Goldstein and N. Andrei, arXiv 1309.7029

\bibitem[3]{key-34}M. Rigol, V. Dunjko, V. Yurovsky and M. Olshanii,
Phys. Rev. Lett. 98, 050405 (2007).

\bibitem[4]{key-32} B. Wouters, M. Brockmann, J. De Nardis, D. Fioretto,
J.-S. Caux arXiv:1405.0172, B. Pozsgay, M. Mestyán, M. A. Werner,
M. Kormos, G. Zaránd, G. Takács arXiv:1405.2843

\bibitem[5]{key-31} M. Takahashi, \textit{Thermodynamics of one-dimensional
solvable models}, (Cambridge University Press, 1999).

\bibitem[6]{key-3}M. Kormos, A. Shashi, Y.-Z. Chou, J.-S. Caux, Adilet
Imambekov Phys. Rev. B \textbf{88}, 205131 (2013) 

\bibitem[7]{key-4} See appendix

\bibitem[8]{key-2}V. E. Korepin, N. M. Bogoliubov and A. G. Izergin,
\textit{Quantum inverse scattering and correlation functions}, (Cambridge
University Press, 1993).

\bibitem[9]{key-33} J. De Nardis, B. Wouters, M. Brockmann and J.-S.
Caux, Phys. Rev. A \textbf{89}, 033601 (2014).

\bibitem[10]{key-1} J. Mossel, J.-S. Caux, J. Phys. A: Math. Theor.
\textbf{45}, 255001, (2012).

\bibitem[11]{key-9} M. Mierzejewski, P. Prelovsek, and T. Prosen
arXiv 1405.2557\end{thebibliography}
\end{document}